# Modeling and Analysis of Power Line Communications for Application in Smart Grid

Moegamat Peck, BSc[1], Genesis Alvarez, BSc[1], Benjamin Coleman, BSc student[1]
Hadis Moradi, PhD candidate[1], Mark Forest, Mentor[2] and Valentine Aalo, Faculty mentor[1]
[1]Florida Atlantic University, Boca Raton, FL, 33431, USA
{mpeck2014, genesisalvar2013, bcoleman2012, hmoradi, aalo}@fau.edu
[2]Florida Power & Light, Boca Raton, FL, 33431, USA mark.forrest@fpl.com

*Abstract–Smart grid is an energy infrastructure that increases energy efficiency by using communication infrastructure, smart meters, smart appliances, automated control and networking, and more. This paper focuses on the Power Line Communication (PLC) aspect and technologies used in the smart grid. There are various challenges and advancements in the smart grid; this research discusses how PLC can improve smart grid performance. In order to provide applicable results, practical PLC system parameters and other required data was obtained from Florida Power and Light (FPL). Modeling of the PLC system with different types of digital modulations was conducted using MATLAB/Simulink software and Python. The benefits and design tradeoffs of Amplitude Shift Keying (ASK), Frequency Shift Keying (FSK), and Phase Shift Keying (PSK) are discussed. The modulation schemes are compared on the basis of their applicability to a practical PLC network by comparing the results of the simulations.*

*Keywords-- Power line communications, Smart grid, Amplitude shift keying, Frequency shift keying, phase shift keying.*

## I. INTRODUCTION

Growing costs of conventional energy with finite resources, Greenhouse Gas (GHG) emissions, climate change issues, security, and reliability of the electric power system have brought many concerns, thus interests toward the development of smart grid and utilizing renewable Distributed Energy Resources (DERs) are increasing all over the world [1],[2].

The generation, transmission and distribution of electric power to consumers is managed and optimized by the smart grid. Power generation via renewable and conventional energy resources is optimized and managed by the smart grid operators at the distribution level in terms of financial and environmental applications [3],[4]. The smart grid includes smart meters, smart appliances, energy efficient resources, and renewable energy resources. Telemetry is used to collect data such as voltage, current flow, power consumption, frequency, phase angles, and temperature. The collection of such data allows power utilities to improve efficiency and sustainability. As power system enhance and integrate with renewable energy resources, the interest in advancement of communication technology increases. The Federal Communications Commission (FCC) has placed certain policies and regulations regarding bandwidth, modulation types, channel coding schemes, operating frequency, and electromagnetic capability limits from fixed indoor/outdoor applications to smart grid applications [14].

In communication technology, there are two types of communication infrastructures; namely wired and wireless technology. The PLC system allows renewable energy integration to sell power to the electric grid [6],[7]. Current challenges for PLC includes the collection of data from alternative clean energy resources to predict future power generation such as wind and solar because of their unpredictable and intermittent nature. Knowledge of daily power generation requirements will decrease the amount of energy wasted in power generation sector. Another challenge that data collection faces is the delay in receiving the required information. As a result, communication system design objective is tom decreasing the measurement time so that different parts of the power grid will interact properly.

This paper focuses on the use of wired communication technologies for the smart grid that utilize the existing power line infrastructure. PLC systems may use either narrowband PLC or broadband PLC. However, due to the characteristics of the communication channel, analog modulation is not suitable for PLC. Therefore, this paper addresses only digital modulation techniques, including Amplitude Shift Keying (ASK), Frequency Shift Keying (FSK) and Phase Shift Keying (PSK). Using substation connection data provided by Florida Power and Light (FPL), the power line channel forming the Atlantic and Yamato substation connection is modeled and considered for data transmission and electrical fault detection. For data transmission, ASK, FSK and PSK communication systems are designed, simulated, and evaluated using MATLAB/Simulink. In this way, a repeatable simulation and design process for PLC data transmission is presented. Electrical fault detection is also demonstrated in a similar fashion.

## II. SMART GRID CONCEPTUAL MODEL

The smart grid is implemented in the electric power system to make the existing grid more efficient, reliable, secure, and capable to incorporate renewable energy sources. Over time, the energy demand has grown, and rapid reduction of energy resources has become a serious threat. Therefore, sustainable clean energy resources, such as wind and solar energy, have gained attention because of their environmentally friendly characteristics [8]-[10]. The smart grid integrates these sources with the generation, transmission, and distribution of the electric utilities. The smart grid is a two-way communication between the customer and electric utility [16]. It is utilized to manage power, reduce outages probability, and inform the consumer of the rate of power they are consuming with particular smart appliances. In electric utilities, the amount of electricity generated should match the amount of electricity consumed in the entire grid to fulfill power balance. Utilities incorporate a demand side management (DSM) program where, when the energy demand exceeds the capacity for power generation, the utility reduces controllable loads such as HVAC units, pool pumps, and hot water heaters. This is known as load shedding which typically occurs during peak usage time [12].

A device called the Phasor Measurement Unit (PMU) measures the magnitude and phase angle of the electrical waves on an electric grid. Phase angles of the bus voltages in real time are measured by synchronizing the measurements from different locations and are captured by high precision Global Positioning System (GPS) clocks [13]. The transmission sector has also integrated a control system architecture referred to as Supervisory Control and Data Acquisition (SCADA), which is a remote control of substations and generation facilities [14]. Some devices that are remote controlled are feeder breakers, capacitor banks, and relays. Future projects within the smart grid in transmission rely on the availability of real time information for renewable energy sources.

## III. PLC STATUS AND ITS STANDARDIZATION

PLC operates by adding a modulating carrier signal to the existing power line, through a narrowband PLC (NB-PLC) or a broadband PLC (BB-PLC). NB-PLC typically operates at a range of 3 kHz to 500 kHz [15]. Although it has a low band data collection, the distance range that is transmitted is up to several kilometers. NB-PLC is available at a low cost and has high reliability through the OFDM modulation technique. Some of the data transmitted by a NB-PLC is the consumption of power in customer side to the utility in real time. A NB-PLC is able to communicate through transformers due to its lower frequencies [15]. In the transmission sector NB-PLC is used to detect broken insulators, short circuits and open /close feeder breakers, real time sag monitoring, remote station surveillance, fault detection, substation automation, and smart metering applications [16].

BB-PLC operates at a range of 1.8MHz to 250MHz and has a much shorter range. It is used for high speed broadband connections such as the internet [16]. Applications that require BB-PLC are HomePlug Green (GP); HomePlug AV PHY; and High Definition (HD)-PLC. Python was used to generate the output signals of BB-PLC and NB-PLC systems. In Fig. 1, a BB–PLC is displayed and in Fig. 2, the output signal of a NB-PLC is shown.

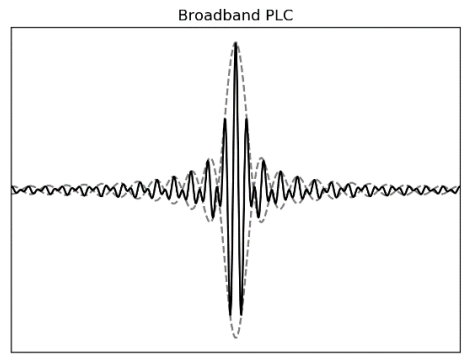

Fig. 1. Broadband PLC

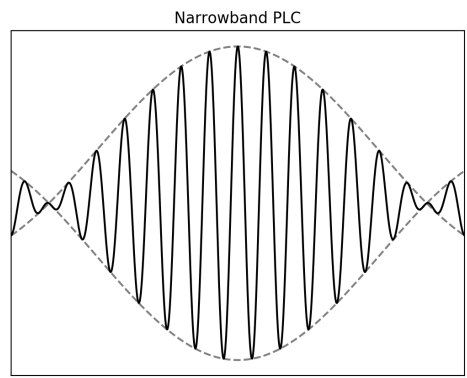

Fig. 2. Narrow band PLC

## IV. APPLICATION TO ATLANTIC-YAMATO SUBSTATION COMMUNICATION LINK

An important component of any smart grid PLC system design is a block diagram and simulation of the system. Here, the power line link between the Atlantic substation and the Yamato substation in the distribution network of Florida Power and Light (FPL) is considered to demonstrate a practical PLC system design flow. The simulation strategies established here have applications in industry and academia, to model the performance of proposed PLC systems.

Two main types of applications are simulated in this research. The first one is data transmission, where information is sent through the power line via digital communications techniques, and the second one is process control, where line faults are detected and circuit breakers are activated by sensing the presence of a carrier signal.

The characteristics of the communication channel were provided by FPL Company. The power line considered here is a 6.07-mile (9800 meter) length of 138 kV high-voltage line between the FPL Atlantic substation and the FPL Yamato substation. For the purposes of modeling the communication channel, the resistance and inductance per meter were approximated as 0.2 ohm and 0.5 mH. The available bandwidth of the channel is restricted between the range from 99 kHz to 400 kHz allocated by FPL for PLC and is selected 301 kHz in this paper. All other system parameters were estimated using established guidelines from data made available by FPL, leading to a Linear Time Invariant (LTI) model of the channel. Although the channel is generally time-varying for PLC systems, due to the difference in power use throughout the day, the LTI model that is applied here represents a worst-case scenario for the transmitted communication signal.

## V. Fault Detection & Simulations

An abnormal electric current is referred to as a fault current. Typical causes of a fault include lightning strikes, insulation contamination, equipment insulation failure, and animal spanning two lines [17].

The two types of faults are unsymmetrical faults and symmetrical faults. A symmetrical fault is also known as a balanced fault and is divided into two types, which are line-to-line-to- ground (L-L-L-G) and line-to-line-to-line (L-L-L). These faults may cause thermal damage to the equipment. The unsymmetrical faults are more common and less severe. It consists of single line–to–ground, line–to–line, double line–to–ground, and balanced three phase faults.

The average Extra High Voltage (EHV) protective equipment is designed to clear faults within 3 cycles [18]. The powerline transmission is divided into zones and a carrier signal is transmitted on the line to detect if there is a fault. In the case of a fault, it trips the corresponding relay within the respective zone. In order to detect the fault location and create a service restoration, Intelligent Electronic Devices (IEDs) are implemented within the substation, which communicate with external IEDs such as switches, reclosers and sectionalizers [18]. Distance relays, also known as impedance relays, operate on the basis of voltage-to-current ratio. The relay is utilized to open the circuit breaker in the event of a fault.

An example of a fault scenario of a three-phase system is displayed in Fig. 3 implemented in MATLAB/Simulink. The system output 3-phase current are displayed in Fig. 4. As shown in the results, a fault occurs at 0.1 seconds that stops normal system operation.

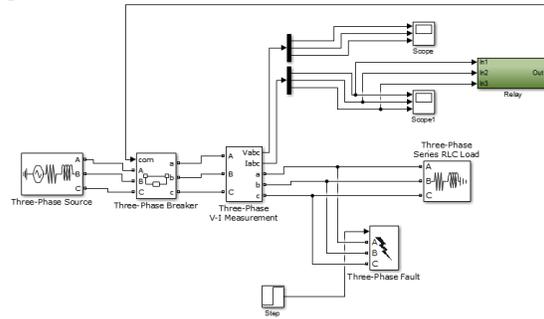

Fig. 3. Phase system Fault analysis

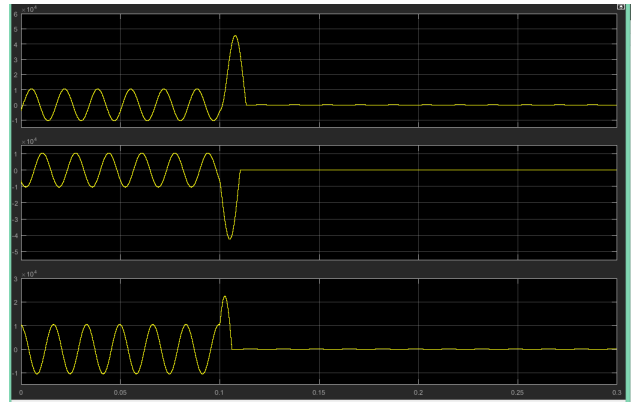

Fig. 4. Current vs. time

## VI. Data Transmission Networks

When binary data is transmitted over a channel, the signal is modulated onto a carrier wave with fixed frequency limits, set by the particular channel [6]. The digital counterparts to the traditional analog modulation schemes of AM, FM, and PM are Amplitude Shift Keying (ASK), Frequency Shift Keying (FSK), and Phase Shift Keying (PSK), respectively. The modulating process consists of amplitude shift keying, frequency shift keying, or phase shift keying demonstrated in Fig. 5. In amplitude shift keying, the amplitude of the carrier is changed depending on the data signal. The high amplitude is represented by bit 1 and the low amplitude is represented by a bit 0. In the demodulation, the carrier is removed from the modulated signal to obtain the original message signal.

Due to the envelope not being constant within the transmitted signal, power amplification becomes inefficient. Three important factors for efficient power line communication consist of the maximum power allowed by the modem that can be transmitted, channel-induced errors, and the color of the background noise [20].

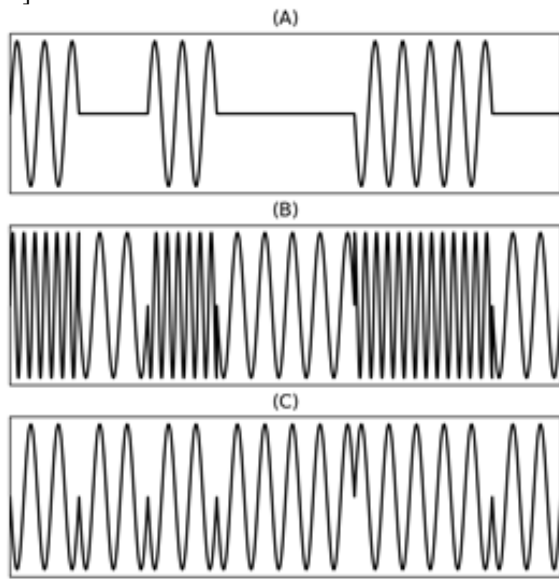

Fig. 5 signaling binary information (a) Amplitude Shift Keying (ASK)(b) Frequency Shift Keying (FSK) (c) Phase Shift Keying (PSK)

Today, smart grid technology such as SCADA, electricity meters, and water meters use the standard distribution automation using the distribution line carrier system known as Frequency Shift Keying (FSK) [21]. FSK consists of two sinusoidal waves with identical amplitudes and distinct frequencies that represent binary bit 1 and 0. The phase in Binary Phase Shift Keying (BPSK) can be varied depending on the message signal. BPSK is robust against interference. BPSK can be utilized to transmit measured data over the power lines on the monitoring system [22]. With the introduction of renewable energy sources to the smart grid, the ability to send data on the amount of generated power by these renewable sources becomes essential. For power line communications, a digital modulation scheme is required to transmit the signals.

## VII. PLC ASK DATA TRANSFER SIMULATION

The major issue with ASK is that it is susceptible to noise interference. For power lines, interference due to noise causes deterioration in system performance. Therefore, ASK is rarely used in PLC. An ASK signal in the PLC network is modeled for a 138kV 60 Hz powerline using a standard ASK architecture. The Fig. 6 shows the simulated ASK model. The results, as shown in Fig. 7, reveal that the interference caused by the channel has a major effect on the ASK system and the output signal.

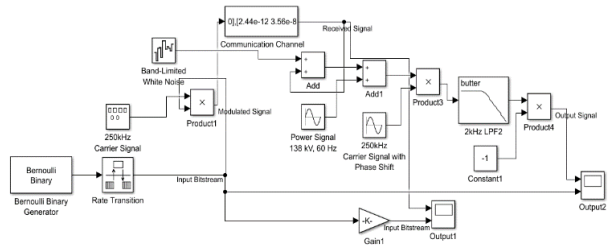

Fig. 6. ASK simulation model

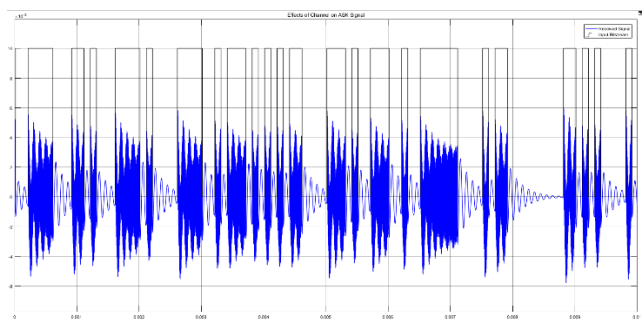

Fig. 7. Effects of channel on ASK signal

## VIII. PLC FSK DATA TRANSFER SIMULATION

On powerlines, FSK is typically used for bit rates of less than 10 kbps, which limits the application to control and signaling and the transmission of small amounts of data [23]. Using conventional FM receiver architecture for the BFSK demodulator, the BFSK transmission was modeled for a 138 kV power line between substations, at exactly 10 kbps with a frequency deviation of 50 kHz and a center frequency of 250 kHz shown in Fig. 8. Although the substation connection is three-phase, enabling multiplexing, only the basic single phase FSK communication system was modeled here as shown in Fig. 9, as the others are duplicates of this system. The transfer function of the powerline was applied to the communication signal, after which the 138 kV power signal was added as represented in Fig. 10.

The bandpass filter had a center frequency of 250 kHz and a bandwidth of 150 kHz. It is also important to note that for the purposes of simulation, rate transition blocks are required by Simulink to properly interface between the digital and analog system components.

The output, presented in Fig. 11, shows that the received signal is identical to the input, except for a time delay and the effect of the impulse response of the filter. The filter impulse response occurs as a result of the suddenly-applied simulation signal. The output shows that FSK transmission and reception was properly modeled and operational for this PLC system.

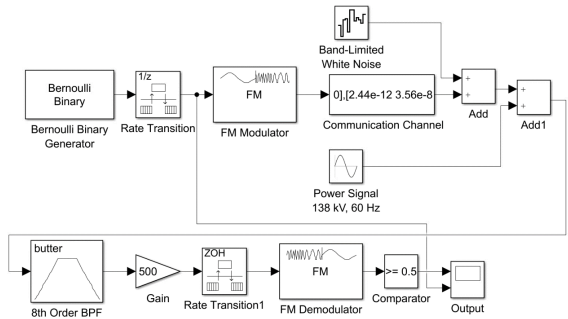

Fig. 8. FSK simulation model

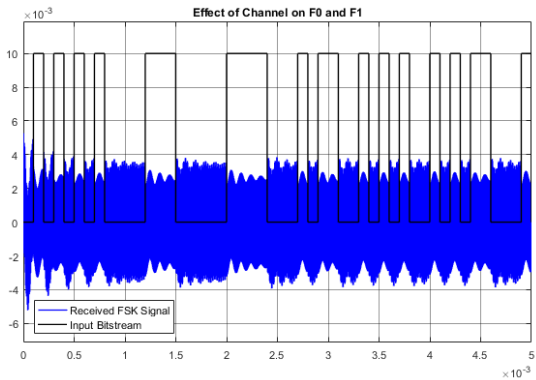

Fig.9 Effects of channel on F0 & F1

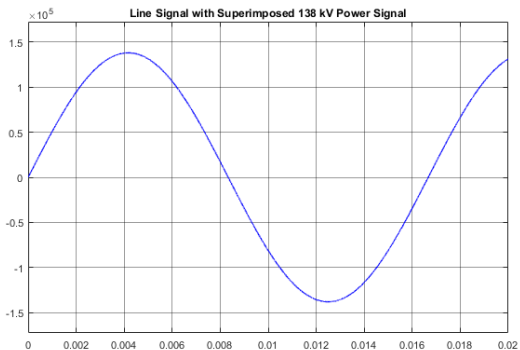

Fig. 10 Power signal

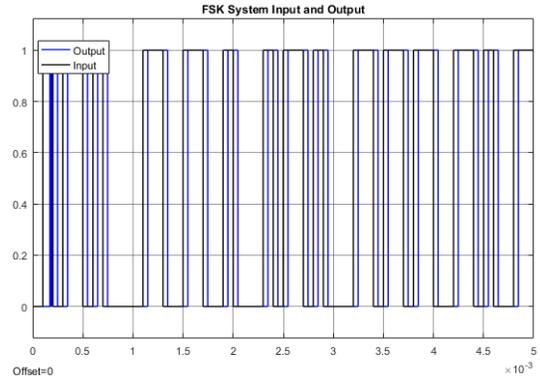

Fig. 11. Input & output response

IX. PLC PSK DATA TRANSFER SIMULATION

PSK was modeled using a standard PSK modulator and demodulator. The PSK signal generated, shown in Fig. 12, depicts the effects that noise has on the modulated signal. This transmission was modeled for a 138 kV, 60 Hz power line, identical to FSK. It has a carrier frequency of 250 kHz. The output of the PSK is similar to the output of the FSK, with a time delay as well as shown in Fig. 13. PSK and FSK are similar but PSK uses less bandwidth. Therefore, more data can be transmitted with the same amount of bandwidth [24].

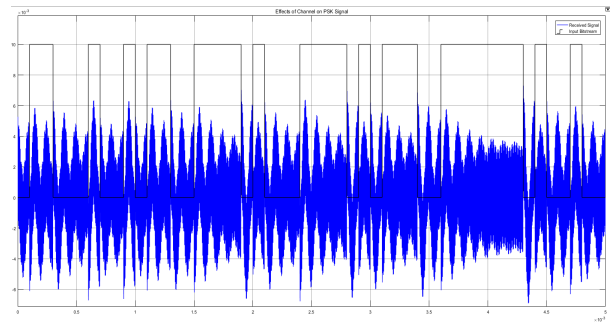

Fig. 12 Effects of channel on PSK

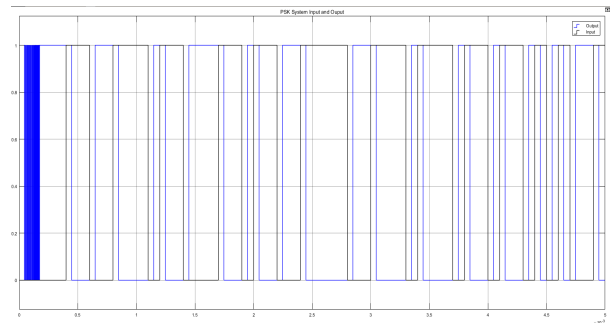

Fig. 13 Input & output response

X. SIMULATION RESULTS

Comparing the results that were collected from each simulation, it is shown that since ASK is highly susceptible to noise, it is not a good fit for PLC applications, particularly as the channel is not intended for communication. PLC systems require robust modulation that can withstand interference adequately. Analysis of the ASK signal in Fig. 6 shows that, due to the noise and channel step response, there is a residual signal when the input is bit 0. This can cause the Bit Error Rate (BER) to increase. To minimize the BER, FSK or PSK should be used. The impulse response for FSK and PSK, shown in Fig. 11 and Fig. 13, respectively, show that the output has minimal to no BER.

XI. Conclusion

In this paper, PLC with different types of digital modulations were simulated and discussed. FSK and PSK were determined to be the best fit for PLC networks. This is due to both being robust modulation that is able to withstand the noise that is present in the system. Noise affects the system in the form of BER. If bits were being misread, the system's performance would diminish. With an FSK or PSK system, this error rate is decreased compared to that of an ASK system. Since ASK modulations varies the amplitude, it is more susceptible to noise, causing a higher BER. Therefore, using an FSK or PSK modulation would increase the performance of the system and, in terms, increase the performance and energy efficiency of the smart grid.